\documentclass{ws-procs9x6}
\usepackage{amsmath}
\usepackage{mathrsfs}
\begin{document}

\hyphenation{ener-gy ave-ra-ge Ave-ra-ge ge-ne-ra-ted re-la-ti-vi-stic dif-fe-ren-ce in-sensi-ti-ve ki-ne-ma-ti-cal Sy-ste-ma-tic sy-ste-ma-tic ave-ra-ging con-ver-ters con-ver-ter la-bo-ra-to-ry se-con-da-ry
se-con-da-ries si-mu-la-tion si-mu-la-tions do-mi-na-te do-mi-na-tes ta-bu-la-tions
Ta-bu-la-tions Di-stri-bu-tions
di-stri-bu-tions di-stri-bu-tion Di-stri-bu-tion Di-spla-ce-ment di-spla-ce-ment Di-spla-ce-ments Bethe power powers ma-te-rial ma-te-rials
di-spla-ce-ments Li-near li-near Ca-sca-de Ca-sca-des ca-sca-de ca-sca-des
ta-bu-la-tion Ta-bu-la-tion re-pe-ti-ti-ve E-lec-tron E-lec-trons
e-lec-tron e-lec-trons De-tec-tion Pro-duc-tion pro-duc-tion
Re-so-lu-tions Re-so-lu-tion re-so-lu-tions re-so-lu-tion
Ope-ra-tion mi-ni-mum Ener-gy Ener-gies fer-ro-mag-net
fer-ro-mag-nets meta-sta-ble meta-sta-bi-lity con-fi-gu-ra-tion
con-fi-gu-ra-tions expo-nen-tially mo-bi-li-ty- mo-bi-li-ties
tem-pe-ra-tu-re tem-pe-ra-tu-res con-cen-tra-tion con-cen-tra-tions
elec-tro-nic elec-tro-nics STMelec-tro-nics sec-tion Sec-tion
Chap-ter chap-ter theo-ry ap-pro-xi-mation ra-dia-tion Ra-dia-tion
ca-pa-ci-tan-ce approaches tran-sport dispersion Ca-lo-ri-me-try
ca-lo-ri-me-try En-vi-ron-ment En-vi-ron-ments en-vi-ron-ment
en-vi-ron-ments Fur-ther-mo-re do-mi-nant ioni-zing pa-ra-me-ter pa-ra-me-ters
O-sa-ka ge-ne-ral exam-ple Exam-ple ca-vi-ty Ca-vi-ty He-lio-sphe-re
he-lio-sphe-re dis-tan-ce Inter-pla-ne-ta-ry inter-pla-ne-ta-ry
ge-ne-ra-li-zed sol-ving pho-to-sphe-re sym-me-tric du-ring
he-lio-gra-phic strea-ming me-cha-nism me-cha-nisms expe-ri-mental
Expe-ri-mental im-me-dia-tely ro-ta-ting na-tu-rally
ir-re-gu-la-ri-ties o-ri-gi-nal con-si-de-red e-li-mi-na-ting
ne-gli-gi-ble stu-died dif-fe-ren-tial mo-du-la-tion ex-pe-ri-ments
ex-pe-ri-ment Ex-pe-ri-ment Phy-si-cal phy-si-cal in-ve-sti-ga-ted
Ano-de Ano-des ano-de ano-des re-fe-ren-ce re-fe-ren-ces
ap-pro-xi-ma-ted ap-pro-xi-ma-te in-co-ming bio-lo-gi-cal
atte-nua-tion other others eva-lua-ted nu-cleon nu-cleons reac-tion
pseu-do-ra-pi-di-ty pseu-do-ra-pi-di-ties esti-ma-ted va-lue va-lues
ac-ti-vi-ty ac-ti-vi-ties bet-ween Bet-ween dis-cre-pan-cy
dis-cre-pan-cies cha-rac-te-ri-stic
cha-rac-te-ri-stics sphe-ri-cally anti-sym-metric ener-gy ener-gies
ri-gi-di-ty ri-gi-di-ties leaving pre-do-mi-nantly dif-fe-rent
po-pu-la-ting acce-le-ra-ted respec-ti-ve-ly sur-roun-ding
sa-tu-ra-tion vol-tage vol-tages da-ma-ge da-ma-ges be-ha-vior
equi-va-lent si-li-con exhi-bit exhi-bits con-duc-ti-vi-ty
con-duc-ti-vi-ties dy-no-de dy-no-des created Fi-gu-re Fi-gu-res
tran-si-stor tran-si-stors Tran-si-stor Tran-si-stors ioni-za-tion
Ioni-za-tion ini-tia-ted sup-pres-sing in-clu-ding maxi-mum mi-ni-mum
vo-lu-me vo-lu-mes tu-ning ple-xi-glas using de-pen-ding re-si-dual har-de-ning li-quid
know-ledge usage me-di-cal par-ti-cu-lar scat-te-ring ca-me-ra se-cond hea-vier hea-vy trans-axial
con-si-de-ration created Hy-po-the-sis hy-po-the-sis usually inte-ra-ction Inte-ra-ction
inte-ra-ctions Inte-ra-ctions pro-ba-bi-li-ty pro-ba-bi-li-ties
fol-low-ing cor-re-spon-ding e-la-stic readers reader pe-riod pe-riods geo-mag-ne-tic sa-ti-sfac-tory ori-gi-nal-ly}

\begin{center}
To appear on the Proceedings of the 13th ICATPP Conference on\\
Astroparticle, Particle, Space Physics and Detectors\\ for Physics Applications,\\ Villa  Olmo (Como, Italy), 3--7 October, 2011, \\to be published by World Scientific (Singapore).
\end{center}
\vspace{-1.5cm}

\title{NUCLEAR AND NON-IONIZING ENERGY-LOSS OF ELECTRONS WITH LOW AND RELATIVISTIC ENERGIES
IN MATERIALS AND SPACE ENVIRONMENT}

\author{M.J. Boschini$^{1,2}$, C. Consolandi$^{*,1}$, M. Gervasi$^{1,3}$, S. Giani$^{4}$, D. Grandi$^{1}$,\\ V. Ivanchenko$^{4}$, P. Nieminem$^{5}$, S. Pensotti$^{3}$, P.G. Rancoita$^{1}$ and M. Tacconi$^{1}$}

\address{$^1$\textit{INFN-Milano Bicocca, P.zza Scienza,3 Milano, Italy}\\
$^2$\textit{CILEA Via R. Sanzio, 4 Segrate, MI-Italy}\\
$^3$\textit{Milano Bicocca University, P.zza della Scienza, 3 Milano, Italy} \\
$^4$\textit{CERN, Geneva, 23, CH-1211, Switzerland}\\
$^5$\textit{ESA, ESTEC, AG Noordwijk (Netherlands) \\}
$^*$E-mail: cristina.consolandi@mib.infn.it}~

\begin{abstract}
The treatment of the electron--nucleus interaction based on the Mott differential cross section was extended to account for effects due to screened Coulomb potentials, finite sizes and finite rest masses of nuclei for electrons above 200\,keV and up to ultra high energies.~This treatment allows one to determine both the total and differential cross sections, thus, subsequently to calculate the resulting nuclear and non-ionizing stopping powers.~Above a few hundreds of MeV, neglecting the effect due to finite rest masses of recoil nuclei the stopping power and NIEL result to be largely underestimated.~While, above a few tens of MeV, the finite size of the nuclear target prevents a further large increase of stopping powers which approach almost constant values.
\end{abstract}

\bodymatter

\section{Introduction}
Nuclei and electrons populate the heliosphere.~Most of the nuclei are galactic cosmic rays (GCR), while electrons can additionally be originated by the Sun and Jupiter's magnetosphere, which is a major source of relativistic electrons in the heliosphere (e.g.,~see Ref.\cite{Meyer_V,Owens2010} and references therein).~Protons and electrons are also major constituents of the Earth's radiation belts.~These particles can interact with materials and onboard electronics in spacecrafts, inducing displacements of atomic nuclei, thus inflicting permanent damages.~As the particle energy increases, for instance above
$\approx 20\,$MeV for protons and $\approx 130\,$\,MeV/nucleon for $\alpha$-particles (e.g.,~see Section~4.2.1.4 and Figure~4.26 at page~418 of Ref.\cite{LR_3rd}), the dominant mechanism for displacement damage is determined by hadronic interactions; for electrons and low-energy nuclei the elastic Coulomb scattering is the relevant physical process to induce permanent damage.~
\par
The non-ionizing energy-loss (NIEL) is the energy lost from particles traversing a unit length of a medium through physical processes resulting in permanent atomic displacements.~The displacement damage is mostly responsible for the degradation of semiconductor devices - like those using silicon - where, for instance, depleted layers are required for normal operation conditions (e.g.~see Ref.\cite{rop_si}).~
The nuclear stopping power and NIEL deposition - due to elastic Coulomb scatterings - from protons, light- and heavy-ions traversing an absorber were previously dealt\cite{Boschini,Boschini_2011} with (see also Sections~1.6, 1.6.1, 2.1.4--2.1.4.2,~4.2.1.6 of Ref.\cite{LR_3rd}).~In the present work, the nuclear stopping power and NIEL deposition due to elastic Coulomb scatterings of electrons are treated up to ultra relativistic energies.
\par
The developed model (i.e.,~see Sects.~\ref{UNscreened_Sect}--\ref{El-En_larger_M}) for screened Coulomb elastic scattering up to relativistic energies is included into Geant4 distribution\cite{geant4} and is available with Geant4 version 9.5 (December 2011).~In Sects.~\ref{El_Nucl_dE/dx},~\ref{El_NIEL}, the nuclear and non-ionizing stopping powers for electrons in materials are treated, while a final discussion is found in Sect.~\ref{Summry_RESULTS}.
\section{Scattering Cross Section of Electrons on Nuclei}
\label{UNscreened_Sect}
The scattering of electrons by unscreened atomic nuclei was treated by Mott\cite{Mott1} (see also Sections~4--4.5 in Chapter~IX of Ref.\cite{Mott_book}) extending a method of Wentzel\cite{Wentzel} (see also Born\cite{Born1}) and including effects related to the spin of electrons\cite{Mott1}\,.~Wentzel's method was dealing with incident and scattered waves on point-like nuclei.~The differential cross section (DCS) -~the so-called \textit{Mott differential cross section} (MDCS) - was expressed by Mott\cite{Mott1} as two conditionally convergent infinite series in terms of Legendre expansions.~In Mott--Wentzel treatment, the scattering occurs on a field of force generating a radially dependent Coulomb - unscreened (screened) in Mott\cite{Mott1} (Wentzel\cite{Wentzel}) - potential.~It has to be remarked that Mott's treatment of collisions of fast electrons with atoms (e.g.,~see Chapter~XVI of~Ref.\cite{Mott_book}) involves the knowledge of the wave function of the atom, thus, in most cases the computation of cross sections depends on the application of numerical methods (see a further discussion in Sect.~\ref{Screened_Sect}).~Furthermore, the MDCS was derived~in the laboratory reference system for infinitely heavy nuclei initially at rest with negligible spin effects and must be numerically evaluated for any specific nuclear target.~Effects related to the recoil and finite rest mass of the target nucleus ($M$) were neglected.~Thus, in this framework the total energy of electrons has to be smaller or much smaller than $M c^2$.
\par
As discussed by Idoeta and Legarda\cite{Idoeta} (e.g.,~see also Refs.\cite{Berger_comp,Fernandez1}), Mott provided an ``exact" differential cross section because no \textit{Born approximation}\footnote[1]{In quantum mechanical potential scattering, the scattered wave may be obtained from the so-called \index{Born!expansion}Born expansion.~The \label{Born_approximation}Born approximation is the first term of the Born expansion (see, for instance, references indicated in Section~1.6.1 Ref.\cite{LR_3rd}).} of any order is employed in its derivation.~Various authors have approximated the MDCS for special situations, usually expressing their results in terms of ratios, $\mathcal{R}$, of the so-obtained approximated differential cross sections with respect to \textit{that one for a Rutherford scattering} (RDCS) - the so-called \index{Rutherford!differential cross section|textbf}\index{Rutherford!formula|textbf}\textit{Rutherford's formula}, see Section~1.6.1 of Ref.\cite{LR_3rd} - for an incoming particle with $z=1$ given by:
\begin{eqnarray}
% \nonumber to remove numbering (before each equation)
  \label{Rutherd_c_s_com_plab}  \frac{d\sigma^{\rm Rut}}{d\Omega}  & = &
 \left(\! \frac{Ze^2}{ p \beta c}\!\right)^2
 \frac{1}{\left(1-\cos \theta \right)^2} =\left(\! \frac{Ze^2}{2\, p \beta c}\!\right)^2 \frac{1}{\sin^4(\theta/2)} \\
  \nonumber &  =&  \left(\!
\frac{Ze^2}{2\, m c^2 \beta^2 \gamma}\!\right)^2
 \frac{1}{ \sin^4(\theta/2)},
\end{eqnarray}
where $m$ is the electron rest mass, $Z$ is the atomic number of the target nucleus, $\beta = v/c$ with $v$ the electron velocity and $c$ the speed of light; $\gamma$ is the corresponding \textit{Lorentz factor}; $p$ and $\theta$ are the momentum and scattering angle of the electron, respectively; finally, since the interaction is isotropic with respect to the azimuthal angle, it is worth noting that $d\Omega$ can be given as
\begin{equation}\label{solid_angle}
d\Omega   = 2\pi \sin\theta \,d\theta.
\end{equation}
The MDCS is usually expressed as:
\begin{equation}\label{MDCS_Mott_R}
   \frac{d\sigma^{\rm Mott}(\theta)}{d\Omega}  = \frac{d\sigma^{\rm Rut}}{d\Omega} \,\, \mathcal{R}^{\rm Mott},
\end{equation}
where $\mathcal{R}^{\rm Mott}$ (as above mentioned) is the ratio between the MDCS and RDCS.~In particular, Bartlett--Watson\cite{Bartlett_Watson} determined cross sections for nuclei with atomic number $Z=80$ and energies from 0.024 up to 1.7\,MeV (see also Ref.\cite{Sherman}).~McKinley and Feshbach\cite{McKinley} expanded Mott's series in terms of power series in $\alpha Z$ (with $\alpha$ the fine-structure constant) and $(\alpha Z)/\beta$; these expansions, which give results accurate to 1\% up to atomic numbers $Z\approx 40$ (e.g.,~see discussions in Refs.\cite{Curr,Cahn}), were further simplified to obtain an approximate analytical formula with that accuracy for $\alpha Z \leq 0.2$.~Feshbach\cite{Feshbach1} tabulated values of the differential cross section as a function of scattering angle for nuclei with atomic number up to 80 and electrons with kinetic energies larger than 4\,MeV.~Curr\cite{Curr} reported values of the differential cross section as a function of scattering angle accurate at 1\% for $(\alpha Z)/\beta \lesssim 0.6$; while Doggett and Spencer\cite{Doddett} tabulated the MDCS for energies from 10 down to 0.05\,MeV.~Recently, Idoeta and Legarda\cite{Idoeta} provided a further series transformations and made a systematic comparison with those from McKinley and Feshbach\cite{McKinley}\,, Curr\cite{Curr}\,, Doggett and Spencer\cite{Doddett}\,.~For electrons with kinetic energies from several keV up to 900\,MeV and target nuclei with $1 \leqslant Z \leqslant 90$, Lijian, Quing and Zhengming\cite{Lijian} provided a practical interpolated expression [Eq.~(\ref{Mott_R_interp_expr})] for $\mathcal{R}^{\rm Mott}$ with an average error less than 1\%; in the present treatment, that expression - discussed in Sect.~\ref{Mott_R_approx} - is the one assumed for $\mathcal{R}^{\rm Mott}$ in Eq.~(\ref{MDCS_Mott_R}) hereafter.
\par
The analytical expression derived by McKinley and Feshbach\cite{McKinley} - mentioned above - for the ratio with respect to Rutherford's formula [Equation~(7) of~Ref.\cite{McKinley}] is given by:
\begin{equation}\label{eq:McF_R}
\mathcal{R}^{\rm McF} = 1-\beta^2\sin^2\!\left(\theta/2\right) +Z\,\alpha\beta\pi \sin\!\left(\theta/2\right)\left[1- \sin\!\left(\theta/2\right)\right]
\end{equation}
with the corresponding differential cross section (McFDCS)
\begin{equation}\label{eq:McF}
\frac{d\sigma^{\rm McF}}{d\Omega}  = \frac{d\sigma^{\rm Rut}}{d\Omega} \,\,\mathcal{R}^{\rm McF},
\end{equation}
where ${d\sigma^{\rm Rut}}/{d\Omega}$ is from Eq.~(\ref{Rutherd_c_s_com_plab}).~It has to be remarked that for positrons, the ratio $\mathcal{R}^{\rm McF}_{\rm pos}$ becomes
\begin{equation}\label{eq:McF_R_pos}
\mathcal{R}^{\rm McF}_{\rm pos} = 1-\beta^2\sin^2\!\left(\theta/2\right) -Z\,\alpha\beta\pi \sin\!\left(\theta/2\right)\left[1- \sin\!\left(\theta/2\right)\right]
\end{equation}
(e.g.,~see Equation~(6) of Ref.\cite{Oen}\,).~Furthermore, for $M c^2$ much larger than the total energy of incoming electron energies the distinction between laboratory
(i.e.,~the system in which the target particle is initially at rest) and center-of-mass (CoM) systems disappears (e.g.,~see discussion in Section~1.6.1 of Ref.\cite{LR_3rd}).~Furthermore, in the CoM of the reaction the energy transferred from an electron to a nucleus initially at rest in the laboratory system (i.e., its recoil kinetic energy $T$) is related to the maximum energy transferable $ T_{\rm max}$ as
\begin{equation}\label{T__rela_thata_prime}
T = T_{\rm max} \,\sin^2
 (\theta'/2)
\end{equation}
[e.g.,~see Equations~(1.27,~1.95) at page~11 and~31, respectively, of~Ref.\cite{LR_3rd}], where $\theta'$ is the scattering angle in the CoM system.~From Eqs.~(\ref{solid_angle},~\ref{T__rela_thata_prime}) one obtains
\begin{equation}\label{dT__rela_thata_prime}
dT =  \frac{ T_{\rm max}}{4 \pi}  \, d\Omega'  .
\end{equation}
Since $\theta$ is $ \approx \theta'$ for $Mc^2$ much larger than the electron energy, one finds that Eq.~(\ref{T__rela_thata_prime}) can be approximated as
\begin{eqnarray}
% \nonumber to remove numbering (before each equation)
\label{T_lab}  T & \simeq & T_{\rm max} \sin^2\!\left(\theta/2\right), \\
\label{T_lab_r} \Longrightarrow \sin^2\!\left(\theta/2\right) &=&  \frac{T}{ T_{\rm max}}
\end{eqnarray}
and
\begin{equation}\label{dT_lab}
dT \simeq \frac{ T_{\rm max}}{4 \pi}  \, d\Omega .
\end{equation}
Using Eqs.~(\ref{eq:McF_R},~\ref{T_lab_r},~\ref{dT_lab}), Eqs.~(\ref{Rutherd_c_s_com_plab},~\ref{eq:McF}) can be respectively rewritten as:
\begin{eqnarray}
% \nonumber to remove numbering (before each equation)
 \nonumber \frac{d\sigma^{\rm Rut}}{d\Omega}  &=&\frac{ T_{\rm max}}{4 \pi} \frac{d\sigma^{\rm Rut}}{dT} \\
\label{Rutherd_c_s_com_plab_T} \Longrightarrow  \frac{d\sigma^{\rm Rut}}{dT}   &=& \left(\! \frac{Ze^2}{ p \beta c}\!\right)^2\frac{ \pi T_{\rm max}}{T^2},\\
 \nonumber \frac{d\sigma^{\rm McF}}{T}  &=  &\left(\! \frac{Ze^2}{ p \beta c}\!\right)^2\frac{ \pi T_{\rm max}}{T^2}\\
 \nonumber & & \times\left[1-\beta^2  \frac{T}{ T_{\rm max}} +Z\,\alpha\beta\pi  \sqrt{\frac{T}{ T_{\rm max}}}\left(1- \sqrt{\frac{T}{ T_{\rm max}}}\right) \right]\\
\label{eq:McF_T} \Longrightarrow \frac{d\sigma^{\rm McF}}{T} \! & =&\!\! \left(\! \frac{Ze^2}{ p \beta c}\!\right)^2\!\frac{ \pi T_{\rm max}}{T^2} \!\!
\left[1\!-\!\beta \frac{T}{ T_{\rm max}}\!  \left( \beta\! + \! Z\alpha\pi \right)\! + \!Z\alpha\beta\pi \! \sqrt{\frac{T}{ T_{\rm max}}} \right]\!\\
 \nonumber & =& \!\! \left(\! \frac{Ze^2}{ p \beta c}\!\right)^2\!\frac{ \pi T_{\rm max}}{T^2} \,\,
\mathcal{R}^{\rm McF}(T)
\end{eqnarray}
with
\begin{equation}\label{R_McF_T}
   \mathcal{R}^{\rm McF}(T) = \left[1\!-\!\beta \frac{T}{ T_{\rm max}}\!  \left( \beta\! + \! Z\alpha\pi \right)\! + \!Z\alpha\beta\pi \! \sqrt{\frac{T}{ T_{\rm max}}} \right]
\end{equation}
[e.g.,~see Equation~(11.4) of~Ref.\cite{Seitz}, see also Ref.\cite{Cahn} and references therein].~Similarly, for positrons one finds
\[
\frac{d\sigma^{\rm McF}_{\rm pos}}{T} \!  =\!\! \left(\! \frac{Ze^2}{ p \beta c}\!\right)^2\!\frac{ \pi T_{\rm max}}{T^2} \!\!
\left[1\!-\!\beta \frac{T}{ T_{\rm max}}\!  \left( \beta\! -\!  Z\alpha\pi \right)\! - \!Z\alpha\beta\pi \! \sqrt{\frac{T}{ T_{\rm max}}} \right]\!
\]
[e.g.,~see Refs.\cite{Cahn,Oen} and references therein].~Finally, in a similar way the MDCS [Eq.~(\ref{MDCS_Mott_R})] is
\begin{eqnarray}
% to remove numbering (before each equation)
\nonumber  \frac{d\sigma^{\rm Mott}(T)}{dT} &=& \frac{d\sigma^{\rm Rut}}{dT} \,\,\mathcal{R}^{\rm Mott}(T)\\
 \label{MDCS_Mott_R_T}  &=& \left(\! \frac{Ze^2}{ p \beta c}\!\right)^2\!\frac{ \pi T_{\rm max}}{T^2}   \,\,          \mathcal{R}^{\rm Mott}(T)
\end{eqnarray}
with $\mathcal{R}^{\rm Mott} (T)$ from Eq.~(\ref{Mott_R_interp_expr_T}).
\subsection{Interpolated Expression for $\mathcal{R}^{\rm Mott}$}
\label{Mott_R_approx}
As mentioned in Sect.~\ref{UNscreened_Sect}, Curr\cite{Curr} derived $ \mathcal{R}^{\rm Mott}$ as a function the atomic number $Z$ of the target nucleus and velocity $\beta c$ of the incoming electron at several scattering angles from $\theta =30^{^{\circ}}$ up to $180^{^{\circ}}$.~Recently, Lijian, Quing and Zhengming\cite{Lijian} provided a practical interpolated expression [Eq.~(\ref{Mott_R_interp_expr})] which is a function of both $\theta$ and $\beta$ for electron energies from several keV up to 900\,MeV,~i.e.,
\begin{equation}\label{Mott_R_interp_expr}
  \mathcal{R}^{\rm Mott} = \sum_{\textrm{j}=0}^4 a_\textrm{j}(Z,\beta)(1-\cos\theta)^{\textrm{j}/2},
\end{equation}
where
\begin{equation}\label{Mott_R_interp_expr_coeff}
a_\textrm{j}(Z,\beta)=\sum_{\textrm{k}=1}^6 b_\textrm{k,j}(Z)(\beta-\overline{\beta})^{\textrm{k}-1},
\end{equation}
and
$\overline{\beta}\,c=0.7181287 \,c$ is the mean velocity of electrons within the above mentioned energy range.~The coefficients $b_\textrm{k,j}(Z)$ are listed in~Table 1 of~Ref.\cite{Lijian} for $1 \leqslant Z \leqslant 90$.~
\par
\begin{figure}[t]
\vspace{-0.7cm}
\begin{center}
%\hspace{-1.cm}
\psfig{file=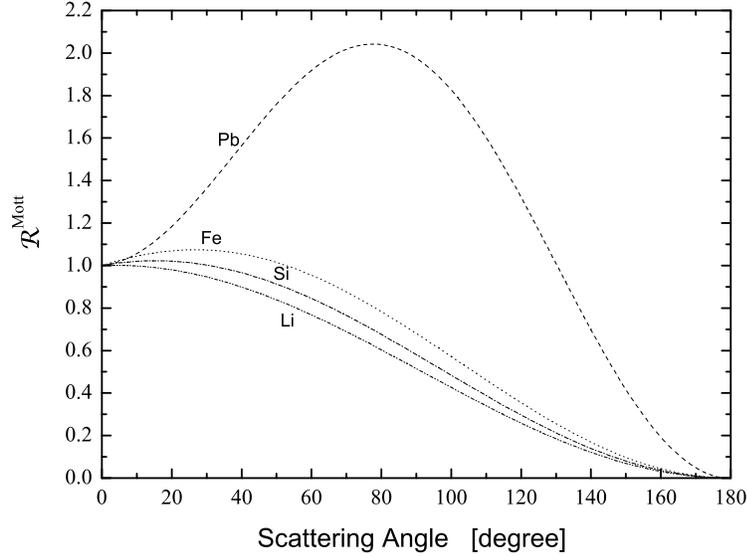,width=4.5in}
\end{center}
\vspace{-0.7cm}
\caption{$\mathcal{R}^{\rm Mott}$ obtained from Eq.~(\ref{Mott_R_interp_expr}) at 100\,MeV for Li, Si, Fe and Pb nuclei as a function of scattering angle.}
\label{Mott_R_Li_china}
\end{figure}

At 10, 100 and 1000\,MeV for Li, Si, Fe and Pb, values of $\mathcal{R}^{\rm Mott}$ were calculated using both Curr\cite{Curr} and Lijian, Quing and Zhengming\cite{Lijian} methods and found to be in a very good agreement.~It has to be remarked that with respect to the values of $\mathcal{R}^{\rm McF} $ obtained from Eq.~(\ref{eq:McF_R}) at 100\,MeV one finds an average variation of about 0.2\%, 3.2\% and 8.8\% for Li, Si and Fe nuclei, respectively.~However, the stopping power determined using Eq.~(\ref{de/dx_nuclear_el_Nucl}) (i.e.,~with $\mathcal{R}^{\rm Mott}$) differs by less than 0.5\% with that calculated using Eq.~(\ref{de/dx_nuclear_el_Nucl_McF}) (i.e.,~with $\mathcal{R}^{\rm McF}$).~$\mathcal{R}^{\rm Mott}$ obtained from Eq.~(\ref{Mott_R_interp_expr}) at 100\,MeV is shown in Fig.~\ref{Mott_R_Li_china} for Li, Si, Fe and Pb nuclei as a function of the scattering angle.~Furthermore, it has to be pointed out that the energy dependence of $\mathcal{R}^{\rm Mott}$ from Eq.~(\ref{Mott_R_interp_expr}) was studied and observed to be negligible above $\approx 10\,$MeV [as expected from Eq.~(\ref{Mott_R_interp_expr_coeff})].
\par
Finally, from Eqs.(\ref{T__rela_thata_prime},~\ref{Mott_R_interp_expr}) [e.g.,~see also Equation~(1.93) at page~31 of~Ref.\cite{LR_3rd}], one finds that $\mathcal{R}^{\rm Mott}$ can be expressed in terms of the transferred energy $T$ as
\begin{equation}\label{Mott_R_interp_expr_T}
  \mathcal{R}^{\rm Mott} (T)=  \sum_{\textrm{j}=0}^4 a_\textrm{j}(Z,\beta)\left(\frac{2T}{T_{\rm max}}\right)^{\textrm{j}/2}.
\end{equation}
\subsection{Screened Coulomb Potentials}
\label{Screened_Sect}
As already mentioned in Sect.~\ref{UNscreened_Sect}, a complete treatment of electron interactions with atoms (e.g.,~see Chapter~XVI of~Ref.\cite{Mott_book}) involves the knowledge of the wave function of the target atom and, thus - as remarked by Fernandez-Vera, Mayol and Salvat\cite{Fernandez1} -, a relevant amount of numerical work when the kinetic energies of electrons exceed a few hundreds of keV.~
\par
The simple scattering model due to Wentzel\cite{Wentzel} - with a single exponential screening function [e.g.,~see Equation~(2.71) at page~95 of~Ref.\cite{LR_3rd}\,, Equation~(21) in~Ref.\cite{Fernandez} and Ref.\cite{Wentzel}] - was repeatedly employed in treating single and multiple Coulomb scattering with screened potentials (e.g,~see~Ref.\cite{Fernandez} - and references therein - for a survey of such a topic and also~Refs.\cite{Boschini,Boschini_2011,Moliere,m_sca_other,Butkevick}).~Neglecting effects like those related to spin and finite size of nuclei, for proton and nucleus interactions with nuclei it was shown that the resulting elastic differential cross section of a projectile with bare nuclear-charge $ez$ on a target with bare nuclear-charge $eZ$ differs from the \index{Rutherford!differential cross section}Rutherford differential cross section (RDCS) by an additional term - the so-called \textit{screening parameter} - which prevents the divergence of the cross section when the angle $\theta$ of scattered particles approaches $0^\circ$ [e.g.,~see~Refs.\cite{Boschini,Boschini_2011,Moliere,m_sca_other,Butkevick} (see also references therein) and Section~1.6.1 of~Ref.\cite{LR_3rd}].~It has to be remarked that the \index{Rutherford!differential cross section}RDCS for $z=1$ particles can also be employed to describe the scattering of non-relativistic electrons with unscreened nuclei (e.g,~see Refs.\cite{Mott1,Idoeta} and references therein).~As derived by Moli\`{e}re\cite{Moliere} for the single Coulomb scattering using a \index{Thomas--Fermi!potential}\textit{Thomas--Fermi potential},~for $z=1$ particles the \index{Screening!parameter}\textit{screening parameter} $A_{\rm s,M}$ [e.g.,~see Equation~(21) of Bethe\cite{m_sca_other}] is expressed as
\begin{equation}\label{eq:As}
A_{\rm s,M}=\left(\frac{\hbar}{2\,p\ a_{\rm TF}}\right)^2\left[1.13+3.76 \times \left(\frac{\alpha Z}{\beta}\right)^2\right]
\end{equation}
where $\alpha$, $c$ and $\hbar$ are the fine-structure constant, speed of light and reduced Planck constant, respectively; $p$ ($\beta c$) is the momentum (velocity) of the incoming particle undergoing the scattering onto a target supposed to be initially at rest -~i.e., in the laboratory system -; $a_{\rm TF}$ is the screening length suggested by Thomas--Fermi~(e.g.,~see Refs.\cite{Thomas,Fermi_TF})
\begin{equation}\label{T_F_sc_rad}
     {\rm {a_{TF}}}=\frac{C_{\rm TF} \,{\rm {a_{0}}} }{Z^{1/3}}
\end{equation}
with \[
{\rm {a_{0}}} =\frac{\hbar^2}{m e^2}
\]
the \index{Bohr!radius}Bohr radius, $m$ the electron rest mass and
\[
C_{\rm TF} = \frac{1}{2}\left(\frac{3 \, \pi}{4} \right)^{2/3} \simeq 0.88534
\]
a constant introduced in the
\index{Thomas--Fermi!model}Thomas--Fermi model [e.g.,~see Equations~(2.73,~2,82) - at page 95 and 99, respectively - of~Ref.\cite{LR_3rd} and Ref.\cite{Boschini_2011}\,, see also references therein].~The modified Rutherford's formula [$ {d\sigma^{\rm WM}(\theta)}/{d\Omega}$],~i.e., the \index{Wentzel--Moli\`{e}re!differential cross section|textbf}\textit{differential cross section} - obtained from the \index{Wentzel--Moli\`{e}re!treatment of the single scattering}Wentzel--Moli\`{e}re treatment of the single scattering on screened nuclear potentials - is given by [e.g.,~see Equation~(2.84) of~Ref.\cite{LR_3rd}\,, Section~2.3 in~Ref.\cite{Fernandez} and Ref.\cite{Boschini_2011} (see also references therein)]:
\begin{eqnarray}
% \nonumber to remove numbering (before each equation)
  \label{eq:diff_cross_W_M_model_sin} \frac{d\sigma^{\rm WM}(\theta)}{d\Omega} &=& \left(\frac{zZe^2}{2\,p \, \beta c }\right)^2\frac{1}{\left[A_{\rm s,M}+\sin^2({\theta}/2)\right]^2} \\
 \nonumber  &=& \frac{d\sigma^{\rm Rut}}{d\Omega} \frac{\sin^4(\theta/2)}{\left[A_{\rm s,M}+\sin^2({\theta}/2)\right]^2} \\
 \label{eq:diff_cross_W_M_model_sin_F_F_screening}  &=& \frac{d\sigma^{\rm Rut}}{d\Omega} \,\,\mathfrak{F}^2(\theta) .
\end{eqnarray}
with
\begin{equation}\label{F_F_screeing}
\mathfrak{F}(\theta) = \frac{\sin^2(\theta/2)}{A_{\rm s,M}+\sin^2({\theta}/2)}.
\end{equation}
$\mathfrak{F}(\theta)$ - the so-called \textit{screening factor} - depends on the scattering angle $\theta$ and screening parameter $A_{\rm s,M}$.~As discussed in Sect.~\ref{El-En_larger_M}, in the DCS the term $A_{\rm s,M}$ cannot be neglected [Eq.~(\ref{eq:diff_cross_W_M_model_sin_F_F_screening})] for scattering angles ($\theta$) within a forward (with respect to the electron direction) angular region narrowing with increasing energy from several degrees (for high-$Z$ material) at 200\,keV down to less than or much less than a mrad above 200\,MeV.
%%%%%%%%%%%%%%%%%%%%%%%%%%%%%%%%%%%%%%%%%%%%%%%%%%%
%%%%%%%%%%%%%%%%%%%%%%%%%%%%%%%%%%%%%%%%%%%%%%%%%%%%%%
\par
An approximated description of elastic interactions of electrons with screened Coulomb fields of nuclei can be obtained factorizing the MDCS,~i.e., involving Rutherford's formula [${d\sigma^{\rm Rut}}/{d\Omega}$] for particles with $z=1$, the screening factor $\mathfrak{F}(\theta)$ and the ratio $\mathcal{R}^{\rm Mott}$ between RDCS and MDCS:
\begin{equation}\label{Factorization_Mott_screened}
   \frac{d\sigma^{\rm Mott}_{\rm sc}(\theta)}{d\Omega}   \simeq \frac{d\sigma^{\rm Rut}}{d\Omega} \,\, \mathfrak{F}^2(\theta)\,\, \mathcal{R}^{\rm Mott}
\end{equation}
[e.g.,~see Equation~(1) of~Ref.\cite{Idoeta}\,, Equation~(A34) at page~208 of~Ref.\cite{Berger_comp}\,, see also Ref.\cite{Fernandez1} and citations from these references].~Thus, the corresponding screened differential cross section derived using the analytical expression from McKinley and Feshbach\cite{McKinley} can be approximated with
\begin{equation}\label{Factorization_McF_screened}
   \frac{d\sigma^{\rm McF}_{\rm sc}(\theta)}{d\Omega}   \simeq \frac{d\sigma^{\rm Rut}}{d\Omega} \,\, \mathfrak{F}^2(\theta)\,\, \mathcal{R}^{\rm McF}.
\end{equation}
It has to be remarked - as derived by Zeitler and Olsen\cite{Zeitler} - that spin and screening effects can be separately treated for small scattering angles; while at large angles (i.e.,~at large momentum transfer), the factorization is well suited under the condition that
\[
2 Z^{4/3} \alpha^2 \frac{1}{\beta^2 \gamma} \ll 1
\]
(e.g.,~see Refs.\cite{Zeitler,Idoeta}).~Zeitler and Olsen\cite{Zeitler} suggested that for electron energies above 200\,keV the overlap of spin and screening effects is small for all elements and for all energies; for lower energies the overlapping of the spin and screening effects may be appreciable for heavy elements and large angles.
\subsection{Finite Nuclear Size}
\label{Finite_Nucl_size}
As suggested by Fernandez-Vera, Mayol and Salvat\cite{Fernandez1}\,, above 10\,MeV the effect of the finite nuclear size has to be taken into account in the treatment of the electron--nucleus elastic scattering.~With increasing energies, deviations from a point-like behavior (see, for instance, Figure~4 of Ref.\cite{Fernandez1}, Ref.\cite{Helm,Ho57} and references therein) were observed at large angles where the screening factor [Eq.~(\ref{F_F_screeing})] is $\approx 1$.~
\par
The ratio between the actual measured and that expected from the point-like differential cross section (e.g.,~the MDCS) expresses the square of the \textit{nuclear form factor} ($\left|F \right|$) which, in turn, depends on the momentum transfer $q$, i.e., that acquired by the target initially at rest:
\begin{equation}\label{momentum_transfer}
q=\frac{\sqrt{T(T+2Mc^2)}}{c},
\end{equation}
with $T$ from Eq.~(\ref{T__rela_thata_prime}) or, for $Mc^2$ larger or much larger than the electron energy, from its approximate expression Eq.~(\ref{T_lab})
[e.g.,~see Equations~(31,~57,~58) of~Ref.\cite{Ho57}\,, Section~3.1.2 of~Ref.\cite{LR_3rd}\,, Refs.\cite{Fernandez1,Butkevick,Helm,DeVries}].~
\par
The factorized differential cross section for elastic interactions of electrons with screened Coulomb fields of nuclei [Eq.~(\ref{Factorization_Mott_screened})] accounting for the effects due to the finite nuclear size is given by:
\begin{eqnarray}
% \nonumber to remove numbering (before each equation)
 \nonumber \frac{d\sigma^{\rm Mott}_{{\rm sc},F}(\theta)}{d\Omega}  &=& \frac{d\sigma^{\rm Mott}_{\rm sc}(\theta)}{d\Omega} \left|F (q)\right| ^2 \\
\label{Factorization_Mott_screened_NFF} & \simeq & \frac{d\sigma^{\rm Rut}}{d\Omega} \,\, \mathfrak{F}^2(\theta)\,\, \mathcal{R}^{\rm Mott} \left|F (q)\right| ^2
\end{eqnarray}
[e.g.,~see Equation~(18) of~Ref.\cite{Fernandez1}\,, Ref.\cite{Butkevick} and also references therein].~Thus, using the analytical expression derived by McKinley and Feshbach\cite{McKinley} [Eq.~(\ref{eq:McF_R})] one obtains the corresponding screened differential cross section [Eq.~(\ref{Factorization_McF_screened})] accounting for the finite nuclear size effects,~i.e.,
\begin{eqnarray}
% \nonumber to remove numbering (before each equation)
 \nonumber \frac{d\sigma^{\rm McF}_{{\rm sc},F}(\theta)}{d\Omega}  &=& \frac{d\sigma^{\rm McF}_{\rm sc}(\theta)}{d\Omega} \left|F (q)\right| ^2 \\
\label{Factorization_McF_screened_NFF} & \simeq & \frac{d\sigma^{\rm Rut}}{d\Omega} \,\, \mathfrak{F}^2(\theta)\,\, \mathcal{R}^{\rm McF} \left|F (q)\right| ^2\\
 \nonumber  & = & \frac{d\sigma^{\rm Rut}}{d\Omega} \,\, \mathfrak{F}^2(\theta) \left|F (q)\right| ^2\\
\label{Factorization_McF_screened_NFF_1} &  &\times \left\{1\!-\!\beta^2\sin^2\!\left(\theta/2\right) +Z\,\alpha\beta\pi \sin\!\left(\theta/2\right)\left[1- \sin\!\left(\theta/2\right)\right]\right\} .
\end{eqnarray}
In terms of kinetic energy, one can respectively rewrite Eqs.~(\ref{Factorization_Mott_screened_NFF},~\ref{Factorization_McF_screened_NFF}) as
\begin{eqnarray}
% \nonumber to remove numbering (before each equation)
 \label{Factorization_Mott_screened_NFF_T}  \frac{d\sigma^{\rm Mott}_{{\rm sc},F}(T)}{dT}  &=&  \frac{d\sigma^{\rm Rut}}{dT} \,\,\mathfrak{F}^2(T) \,\,\mathcal{R}^{\rm Mott}(T)\,\,\left|F (q)\right| ^2
 \\
\label{Factorization_McF_screened_NFF_T} \frac{d\sigma^{\rm McF}_{{\rm sc},F}(T)}{dT} &\simeq & \frac{d\sigma^{\rm Rut}(T)}{dT} \,\, \mathfrak{F}^2(T)\,\,\, \mathcal{R}^{\rm McF} (T)\,\, \left|F (q)\right| ^2
\end{eqnarray}
with ${d\sigma^{\rm Rut}}/{dT}$ from Eq.~(\ref{Rutherd_c_s_com_plab_T}), $\mathcal{R}^{\rm Mott}(T)$ from Eq.~(\ref{Mott_R_interp_expr_T}), $\mathcal{R}^{\rm McF} (T)$ from Eq.~(\ref{R_McF_T}) and, using  Eqs.~(\ref{T__rela_thata_prime},~\ref{T_lab},~\ref{F_F_screeing}),
\[
\mathfrak{F}(T)= \frac{T}{ T_{\rm max} A_{\rm s,M}+T}.
\]
\par
The nuclear form factor accounts for the spatial distribution of charge density probed in the electron--nucleus scattering [e.g.,~see Equation~(58) of~Ref.\cite{Ho57}\,, Section~3.1.2 of~Ref.\cite{LR_3rd}\,, Refs.\cite{Fernandez1,Butkevick,Helm,DeVries} and references therein].~For instance, among those spherically symmetric treated in literature, one finds that for i) an \textit{exponential charge distribution} ($F_{\rm exp}$) [e.g.,~see Equation~(6) of~Ref.\cite{Butkevick}\,, Equation~(93) at page~252 of~Ref.\cite{Ho57} and references therein], ii) a \textit{Gaussian charge distribution} ($F_{\rm gau}$) [e.g.,~see Equation~(6) of~Ref.\cite{Butkevick} and references therein] and iii) an \textit{uniform--uniform folded charge distribution} over spheres with different radii ($F_{\rm u}$) [e.g.,~see Equation~(22) of~Ref.\cite{Fernandez1}\,, Ref.\cite{Helm} and references therein].~For instance, the form factor $F_{\rm exp}$ is
\begin{equation}\label{NFF_exp}
F_{\rm exp}(q)=\left[1+ \frac{1}{12}\left(\frac{q r_{\rm n}}{\hbar}\right)^2\right]^{-2},
\end{equation}
where $r_{\rm n}$ is the nuclear radius [e.g.,~see Equation~(6) of~Ref.\cite{Butkevick}].~To a first approximation, $r_{\rm n}$ can be parameterized by
\begin{equation}\label{r_n_NFF_exp}
r_{\rm n}= 1.27 A^{0.27}~ \textrm{fm}
\end{equation}
with $A$ the atomic weight [e.g.,~see Equation~(7) of~Ref.\cite{Butkevick}].~Equation~(\ref{r_n_NFF_exp}) provides values of $r_{\rm n}$ in agreement up to heavy nuclei (like Pb and U) with those available, for instance, in Table~1 of~Ref.\cite{DeVries}\,.~The nuclear form factor is 1 for $q=0$ and rapidly decreases with increasing $q$ [e.g.,~see Eq.~(\ref{NFF_exp}), Equation~(6) of~Ref.\cite{Butkevick} and Equation~(22) of~Ref.\cite{Fernandez1} for $F_{\rm exp}$, $F_{\rm gau}$ and $F_{\rm u}$, respectively].~Furthermore, from inspection of Eqs.~(\ref{T__rela_thata_prime},~\ref{T_lab},~\ref{momentum_transfer}) small $q$ are those corresponding to scattering angles within the forward (with respect to the electron direction) angular region which, in turn, narrows with increasing electron energy.~For instance, in lithium the square values ($\left|F (q)\right| ^2$) of these form factors are in agreement within 1\% up to $\theta' \lesssim 124.1^{\circ}$ ($2.4^{\circ}$) at 20\,MeV (1\,GeV); in silicon up to $\theta' \lesssim 138.4^{\circ}$ ($2.4^{\circ}$) at 20\,MeV (1\,GeV); in iron up to $\theta' \lesssim 108.0^{\circ}$ ($2.1^{\circ}$) at 20\,MeV (1\,GeV).~However, as discussed in Sect.~\ref{El-En_larger_M}, these upper angles are larger or much larger with respect to those required to obtain 99\% of the total cross section.~Thus, the usage of any of the above mentioned nuclear form factors -~e.g., $F_{\rm exp}$ as in the present treatment - is expected to be appropriate in the treatment of the transport of electrons in matter, when single scattering mechanisms are relevant, for instance in dealing with the nuclear stopping power and non-ionization energy-loss deposition.~
\subsection{Finite Rest Mass of Target Nucleus}
\label{El-En_larger_M}
The DCS treated in Sects.~\ref{UNscreened_Sect}--\ref{Finite_Nucl_size} is based on the extension of the MDCS %- or its approximated expression from McKinley and Feshbach\cite{McKinley} (McFDCS) -
to include effects due to interactions on screened Coulomb potentials of nuclei and their finite size.~However, in the treatment, the electron energies were assumed to be small (or much smaller) with respect to that ($M c^2$) corresponding to the rest mass ($M$) of target nuclei.~
%%%%%%%%%%%%%%%%%%%%%%%%%%%%%%%%%%%%%%%%%%%%%%%%%%%
\begin{figure}[t]
\vspace{-0.7cm}
\begin{center}
%\hspace{-1.cm}
\mbox{\hspace{-0.3cm}\psfig{file=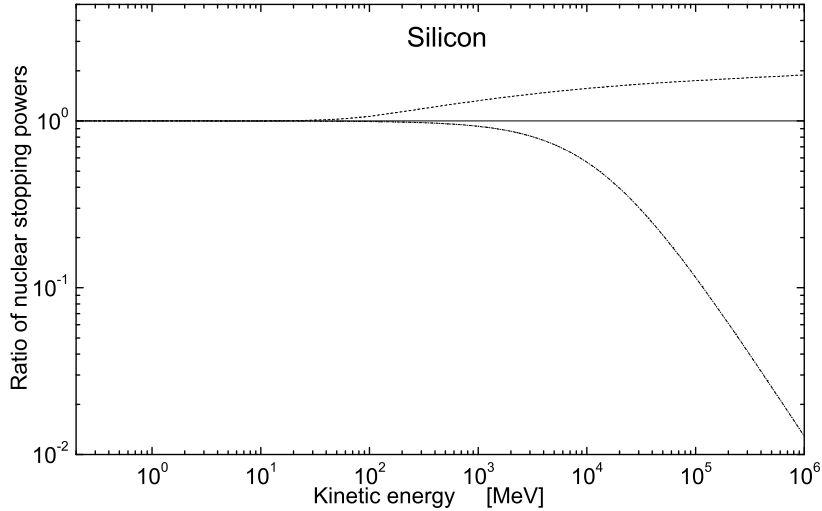,width=4.8in}}
\end{center}
\vspace{-0.7cm}
\caption{As a function of the kinetic energy of electrons from 200\,keV up to 1\,TeV, ratios of nuclear stopping power of electrons in silicon calculated neglecting i) nuclear size effects (i.e., for $\left|F_{\rm exp}\right|^2=1$) (dashed curve) and ii) effects due to the finite rest mass of the target nucleus (dashed and dotted curve) [i.e., in Eq.~(\ref{de/dx_nuclear_el_Nucl_McF}) replacing ${d\sigma^{\rm McF}_{{\rm sc},F,\rm CoM}(T)}/{dT}$ with ${d\sigma^{\rm McF}_{{\rm sc},F}(T)}/{dT}$ from Eq.~(\ref{Factorization_McF_screened_NFF_T})] both divided by that one obtained using Eq.~(\ref{de/dx_nuclear_el_Nucl_McF}).}
\label{fig:RatiodEdx}
\end{figure}
%%%%%%%%%%%%%%%%%%%%%%%%%%%%%%%%%%%%%%%%%%%%%%%%%%%%%%
\par
The Rutherford scattering on screened Coulomb fields - i.e., under the action of a central force - by massive charged particles at energies larger or much larger than $M c^2$ was treated by Boschini et al.\cite{Boschini,Boschini_2011} in the CoM system (e.g.,~see also Sections~1.6,~1.6.1,~2.1.4.2 of~Ref.\cite{LR_3rd} and references therein).~It was shown that the differential cross section [$d\sigma^{\rm WM}(\theta')/d\Omega'$ with $\theta'$ the scattering angle in the CoM system] is that one derived for describing the interaction on a fixed scattering center of a particle with i) momentum $p_{\rm r}'$ equal to the momentum of the incoming particle (i.e.,~the electron in the present treatment) in the CoM system and ii) rest mass equal to the \textit{relativistic reduced mass} $\mu_{\rm rel}$ [e.g.,~see Equations~(1.80,~1.81) at page~28 of~Ref.\cite{LR_3rd}].~$\mu_{\rm rel}$ is given by
\begin{eqnarray}
% \nonumber to remove numbering (before each equation)
\label{reduced_mass_rel}   \mu_{\rm rel} &=&   \frac{m M }{M_{1,2}}\\
\label{reduced_mass_rel1} &=& \frac{m M c}{\sqrt{m^2 c^2 + M^2 c^2 + 2\,M \sqrt{m^2 c^4 + p^2 c^2}}}  ,
\end{eqnarray}
where $p$ is the momentum of the incoming particle (the electron in the present treatment) in the laboratory system: $m$ is the rest mass of the incoming particle (i.e.,~the electron rest mass); finally, $M_{1,2}$ is the invariant mass -~e.g., Section~1.3.2 of~Ref.\cite{LR_3rd} - of the two-particle system.~Thus, the velocity of the interacting particle is
\begin{equation}\label{reduced_mass_beta}
    \beta_{\rm r}' c = c \sqrt{\left[1+ \left(\frac{\mu_{\rm rel} c}{p_{\rm r}' }  \right)^2 \right]^{-1}}
\end{equation}
[e.g.,~see Equation~(1.82) at page~29 of~Ref.\cite{LR_3rd}].~For an incoming particle with $z=1$, $d\sigma^{\rm WM}(\theta')/d\Omega'$ is given by
\begin{equation}\label{eq:diff_cross_W_M_model_r}
   \frac{d\sigma^{\rm WM'}(\theta')}{d\Omega'}  =  \left(\frac{Ze^2}{2\,p_{\rm r}' \, \beta_{\rm r}' c }\right)^2\frac{1}{\left[A_{\rm s}+\sin^2({\theta'}/2)\right]^2},
\end{equation}
with
\begin{equation}\label{eq:As_r_z_1}
A_{\rm s}=\left(\frac{\hbar}{2\,p_{\rm r}'\, a_{\rm TF}}\right)^2\left[1.13+3.76 \times \left(\frac{\alpha Z}{\beta_{\rm r}'}\right)^2\right]
\end{equation}
the screening factor [e.g.,~see Equations~(2.87,~2.88) at page~103 of~Ref.\cite{LR_3rd}].~Equation~(\ref{eq:diff_cross_W_M_model_r}) can be rewritten as
\begin{equation}\label{eq:diff_cross_W_M_model_r_SR_factorized}
   \frac{d\sigma^{\rm WM'}(\theta')}{d\Omega'}  = \frac{d\sigma^{\rm Rut'}(\theta')}{d\Omega'} \,\, \mathfrak{F}_{\rm CoM}^2(\theta')
\end{equation}
with
\begin{equation}\label{Rutherd_c_s_com_red_mass}
\frac{d\sigma^{\rm Rut'}(\theta')}{d\Omega'} =  \left(\!
\frac{Ze^2}{2 p_{\rm r}' \beta_{\rm r}' c }\right)^2
 \frac{1}{  \sin^4(\theta'/2)}
\end{equation}
the corresponding RDCS for the reaction in the CoM system [e.g.,~see Equation~(1.79) at page~28 of~Ref.\cite{LR_3rd}] and
\begin{equation}\label{Wentzel_S_F}
\mathfrak{F}_{\rm CoM}(\theta') = \frac{\sin^2(\theta'/2)}{A_{\rm s}+\sin^2({\theta'}/2)}
\end{equation}
the screening factor.~Using, Eqs.~(\ref{T__rela_thata_prime},~\ref{dT__rela_thata_prime}), one can respectively rewrite Eqs.~(\ref{Rutherd_c_s_com_red_mass},~\ref{Wentzel_S_F},~\ref{eq:diff_cross_W_M_model_r_SR_factorized},~\ref{eq:diff_cross_W_M_model_r}) as
\begin{eqnarray}
% \nonumber to remove numbering (before each equation)
 \label{Rutherd_c_s_com_red_mass_T} \frac{d\sigma^{\rm Rut'}}{dT}  &=&\pi \left(\!
\frac{Ze^2}{ p_{\rm r}' \beta_{\rm r}' c }\right)^2
 \frac{T_{\rm max}}{  T^2} \\\
 \label{Wentzel_S_F_T} \mathfrak{F}_{\rm CoM}(T) &=& \frac{T}{ T_{\rm max}A_{\rm s}+T} \\
 \label{eq:diff_cross_W_M_model_r_SR_factorized_T}  \frac{d\sigma^{\rm WM'}(T)}{dT}   &=& \frac{d\sigma^{\rm Rut'}}{dT} \,\,\mathfrak{F}_{\rm CoM}(T)\\
 \label{eq:diff_cross_W_M_model_r_T}  \frac{d\sigma^{\rm WM'}(T)}{dT}  &=& \pi \left(\!
\frac{Ze^2}{ p_{\rm r}' \beta_{\rm r}' c }\right)^2
 \frac{T_{\rm max}}{ \left(  T_{\rm max}A_{\rm s}+T\right)^2}
\end{eqnarray}
[e.g.,~see Equation~(2.90) at page~103 of Ref.\cite{LR_3rd} or Equation~(13) of~Ref.\cite{Boschini_2011}].
\par
As already mentioned (Sect.~\ref{Screened_Sect}), the screening parameter $A_{\rm s}$ prevents the DCS to diverge -~see last term in Eq.~(\ref{eq:diff_cross_W_M_model_r})~-, i.e.,~for $\theta'$ of the order of or smaller than
\[
\theta_{\rm sc}' = \arcsin \left( 2 \sqrt{A_{\rm s}}\right)
\]
effects due to screening of the nuclear Coulomb field have to be accounted for.~$\theta_{\rm sc}'$  rapidly decreases with increasing the kinetic energies of electrons.~For instance, in iron $\theta_{\rm sc}'$ is $\approx 1.7^\circ$ (0.03\,rad) at 200\,keV and $\approx 0.004^\circ$ ($7.0 \!\times \! 10^{-2}$\,mrad) at 200\,MeV; in silicon, it is $\approx 1.3^\circ$ (0.022\,rad) at 200\,keV and $\approx 0.003^\circ$ ($5.5 \!\times \! 10^{-2}$\,mrad) at 200\,MeV; while, in lithium, it is $\approx 0.75^\circ$ (13\,mrad) at 200\,keV and $\approx 0.002^\circ$ ($3.3 \!\times \! 10^{-2}$\,mrad) at 200\,MeV.~Therefore, in Eq.~(\ref{eq:diff_cross_W_M_model_r_SR_factorized}) the term $A_{\rm s}$ (i.e.,~the screening parameter [Eq.~(\ref{eq:As_r_z_1})]) cannot be neglected for scattering angles within a forward angular region narrowing with increasing energies from a few degrees (for low-$Z$ material) at about 200\,keV down to less than or much less than a mrad above 200\,MeV.~It is worthwhile to remark that in silicon, for instance, $\theta'$ can be approximated with $\theta$ up to a few hundred MeV.
\par
To account for the finite rest mass of target nuclei, the factorized MDCS [Eq.~(\ref{Factorization_Mott_screened_NFF})] has to be re-expressed in the CoM system as:
\begin{eqnarray}
% \nonumber to remove numbering (before each equation)
 \nonumber \frac{d\sigma^{\rm Mott}_{{\rm sc},F,\rm CoM}(\theta')}{d\Omega'} \!\! &=& \!\!\frac{d\sigma^{\rm Mott}_{\rm sc}(\theta',\beta_{\rm r}',p_{\rm r}')}{d\Omega'} \left|F (q)\right| ^2 \\
\nonumber\!\! & \simeq & \!\!\frac{d\sigma^{\rm WM'}(\theta')}{d\Omega'}\, \mathcal{R}^{\rm Mott}_{\rm CoM}(\theta')\,\,\left|F (q)\right| ^2 \!\\
\label{Factorization_Mott_screened_NFF_CoM}\!\! & \simeq & \!\!\frac{d\sigma^{\rm Rut'}(\theta')}{d\Omega'} \,\, \mathfrak{F}_{\rm CoM}^2(\theta')\,\, \mathcal{R}^{\rm Mott}_{\rm CoM}(\theta')\,\,\left|F (q)\right| ^2 \!,
\end{eqnarray}
where $F (q)$ is the nuclear form factor (Sect.~\ref{Finite_Nucl_size}) with $q$ the momentum transfer to the recoil nucleus [Eq.~(\ref{momentum_transfer})]; finally, as discussed in Sect.~\ref{Mott_R_approx}, $\mathcal{R}^{\rm Mott}$ exhibits almost no dependence on electron energy above $\approx 10\,$MeV, thus, since at low energies $\theta \backsimeq \theta'$ and $\beta \backsimeq  \beta_{\rm r}'$, $\mathcal{R}^{\rm Mott}_{\rm CoM}(\theta')$ is obtained replacing $\theta$ and $\beta_{\rm r}'$ with $\theta'$ and $\beta_{\rm r}'$, respectively, in Eq.~(\ref{Mott_R_interp_expr}).
\par
Using the analytical expression derived by McKinley and Feshbach\cite{McKinley}\,, one finds that the corresponding screened differential cross section accounting for the finite nuclear size effects [Eqs.~(\ref{Factorization_McF_screened_NFF},~\ref{Factorization_McF_screened_NFF_1})] can be re-expressed as
\begin{equation}\label{Factorization_McF_screened_NFF_CoM}
% \nonumber to remove numbering (before each equation)
 \frac{d\sigma^{\rm McF}_{{\rm sc},F, \rm CoM}(\theta')}{d\Omega'}  \simeq  \frac{d\sigma^{\rm Rut'}(\theta')}{d\Omega'} \,\, \mathfrak{F}_{\rm CoM}^2(\theta')\,\,\, \mathcal{R}_{\rm CoM}^{\rm McF} (\theta')\,\, \left|F (q)\right| ^2\\
\end{equation}
with
\begin{equation}\label{Factorization_McF_screened_NFF_1_CoM}
\mathcal{R}^{\rm McF}_{\rm CoM} (\theta') =
\left\{\!1\!-\!\beta_{\rm r}^2\sin^2\!\left(\theta'/2\right) \!+\!Z\,\alpha\beta_{\rm r}'\pi \sin\!\left(\theta'/2\right)\left[1\!- \!\sin\!\left(\theta'/2\right)\right]\!\right\} .
\end{equation}
It has to be remarked that scattered electrons are mostly found in the forward or very forward direction.~For instance, using Eq.~(\ref{Factorization_McF_screened_NFF_1_CoM}) one can derive that in lithium $\approx 99\%$ of electrons are scattered with $\theta' \lesssim 0.27^{\circ}$ ($0.007^{\circ}$) at 20\,MeV (1\,GeV); in silicon  with $\theta' \lesssim 0.46^{\circ}$ ($0.009^{\circ}$) at 20\,MeV (1\,GeV); in iron with $\theta' \lesssim 0.6^{\circ}$ ($0.013^{\circ}$) at 20\,MeV (1\,GeV).~
%%%%%%%%%%%%%%%%%%%%%%%%%%%%%%%%%%%%%%%%%%%
\begin{figure}[t]
\vspace{-0.7cm}
\begin{center}
\mbox{\hspace{-0.3cm}\psfig{file=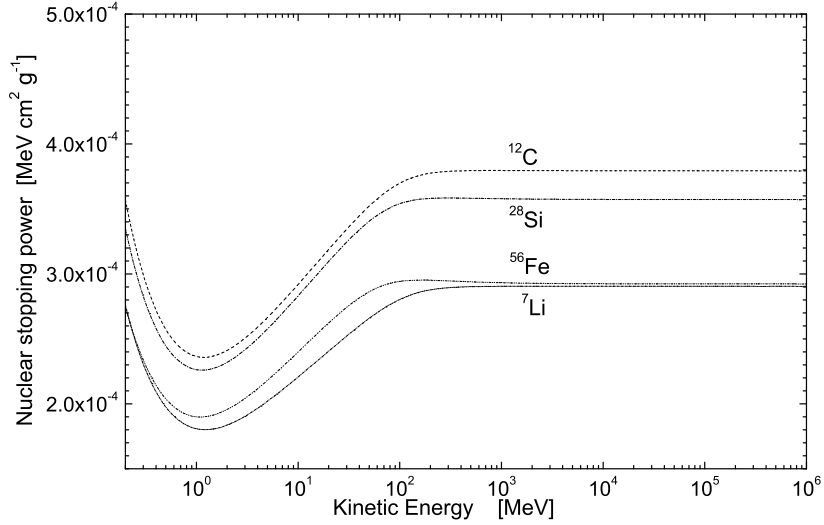,width=4.8in}}
%\psfig{file=dEdx_McF_el_nuc.eps,width=4.5in}
%\vspace{-0.5cm}
%\psfig{file=dEdxMeVcmnew.eps,width=4in}
\end{center}
\vspace{-0.7cm}
\caption{Nuclear stopping powers (in MeV\,cm$^2$/g) in $^7$Li, $^{12}$C, $^{28}$Si and $^{56}$Fe - calculated from Eq.~(\ref{de/dx_nuclear_el_Nucl_McF}) - and divided by the density of the material as a function of the kinetic energy of electrons from 200\,keV up to 1\,TeV.}
\label{fig:el_dEdx}
\end{figure}
%%%%%%%%%%%%%%%%%%%%%%%%%%%%%%%%%%%%%%%%%%%%%%
\par
In terms of kinetic energy $T$, from Eqs.~(\ref{T__rela_thata_prime},~\ref{dT__rela_thata_prime})
one can respectively rewrite Eqs.~(\ref{Factorization_Mott_screened_NFF_CoM},~\ref{Factorization_McF_screened_NFF_CoM}) as
\begin{eqnarray}
% \nonumber to remove numbering (before each equation)
 \label{Factorization_Mott_screened_NFF_CoM_T}  \frac{d\sigma^{\rm Mott}_{{\rm sc},F,\rm CoM}(T)}{dT}  &=&  \frac{d\sigma^{\rm Rut'}}{dT} \,\,\mathfrak{F}_{\rm CoM}^2(T) \,\,\mathcal{R}^{\rm Mott}_{\rm CoM}(T)\,\,\left|F (q)\right| ^2
 \\
\label{Factorization_McF_screened_NFF_CoM_T} \frac{d\sigma^{\rm McF}_{{\rm sc},F, \rm CoM}(T)}{dT} &\simeq & \frac{d\sigma^{\rm Rut'}(T)}{dT} \,\, \mathfrak{F}_{\rm CoM}^2(T)\,\,\, \mathcal{R}_{\rm CoM}^{\rm McF} (T)\,\, \left|F (q)\right| ^2
\end{eqnarray}
with ${d\sigma^{\rm Rut'}}/{dT}$ from Eq.~(\ref{Rutherd_c_s_com_red_mass_T}), $\mathfrak{F}_{\rm CoM}(T)$ from Eq.~(\ref{Wentzel_S_F_T}) and $\mathcal{R}^{\rm McF}_{\rm CoM}(T)$ replacing $\beta$ with $\beta_{\rm r}'$ in Eq.~(\ref{R_McF_T}), i.e.,
\begin{equation}\label{mm}
\mathcal{R}^{\rm McF}_{\rm CoM}(T)=  \left[1\!-\!\beta_{\rm r}'\frac{T}{ T_{\rm max}}\!  \left( \beta_{\rm r}'\! + \! Z\alpha\pi \right)\! + \!Z\alpha\beta_{\rm r}'\pi \! \sqrt{\frac{T}{ T_{\rm max}}} \right].
\end{equation}
Finally, as discussed in Sect.~\ref{Mott_R_approx}, $\mathcal{R}^{\rm Mott}(T)$ exhibits almost no dependence on electron energy above $\approx 10\,$MeV, thus, since at low energies $\theta \backsimeq \theta'$ and $\beta \backsimeq  \beta_{\rm r}'$, $\mathcal{R}^{\rm Mott}_{\rm CoM}(T)$ is obtained replacing $\beta$ with $\beta_{\rm r}'$ in Eq.~(\ref{Mott_R_interp_expr_T}).
%%%%%%%%%%%%%%%%%%%%%%%%%%%%%%%%%%%%%%%%%%%%%%%%%%%%%%%%%%%
\begin{figure}[t]
\vspace{-0.7cm}
\begin{center}
\mbox{\hspace{-0.3cm}
\psfig{file=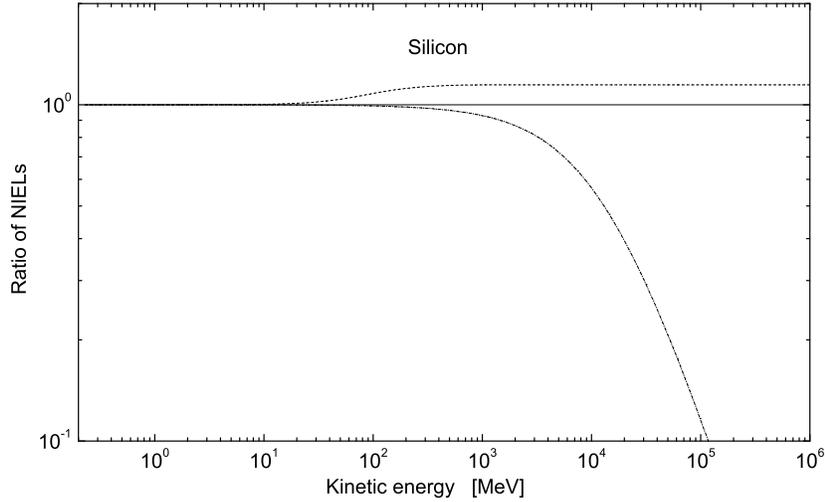,width=4.8in}}
\end{center}
\vspace{-0.7cm}
\caption{For $T_d = 21\,$eV, ratios of NIELs of electrons in silicon calculated as a function of the kinetic energy of electrons from 220\,keV up to 1\,TeV neglecting i) nuclear size effects (i.e., for $\left|F_{\rm exp}\right|^2=1$) (dashed curve) and ii) effects due to the finite rest mass of the target nucleus (dashed and dotted curve) [i.e., in Eq.~(\ref{eq:NIEL_McF}) replacing ${d\sigma^{\rm McF}_{{\rm sc},F,\rm CoM}(T)}/{dT}$ with ${d\sigma^{\rm McF}_{{\rm sc},F}(T)}/{dT}$ from Eq.~(\ref{Factorization_McF_screened_NFF_T})] both divided by that one obtained using Eq.~(\ref{eq:NIEL_McF}).}
\label{fig:RatioNIEL}
\end{figure}
%%%%%%%%%%%%%%%%%%%%%%%%%%%%%%%%%%%%%%%%%%%%%%%%%%%%%%%%%%%%
%
\section{Nuclear Stopping Power of Electrons}
\label{El_Nucl_dE/dx}
Using Eq.~(\ref{Factorization_Mott_screened_NFF_CoM_T}), the nuclear stopping power - in MeV\,cm$^{-1}$ - of Coulomb electron--nucleus interaction can be obtained as
\begin{equation}\label{de/dx_nuclear_el_Nucl}
- \left(\frac{dE}{dx}\right)_{\rm nucl}^{\rm Mott} =  {n_A}\int_{0}^{T_{\rm max}} \frac{d\sigma^{\rm Mott}_{{\rm sc},F,\rm CoM}(T)}{dT} \, T \,d T
\end{equation}
with $n_A$ the number of nuclei (atoms) per unit of volume [e.g., see Equation~(1.71) of Ref.\cite{LR_3rd}] and, finally, the negative sign indicates that energy is lost by electrons (thus, achieved by recoil targets).~Using the analytical approximation derived by McKinley and Feshbach\cite{McKinley},~i.e.,~Eq.~(\ref{Factorization_McF_screened_NFF_CoM_T}), for the nuclear stopping power one finds
\begin{equation}\label{de/dx_nuclear_el_Nucl_McF}
- \left(\frac{dE}{dx}\right)_{\rm nucl}^{\rm McF} =  {n_A}\int_{0}^{T_{\rm max}} \frac{d\sigma^{\rm McF}_{{\rm sc},F,\rm CoM}(T)}{dT} \, T \,d T.
\end{equation}
\par
As already mentioned in Sect.~\ref{El-En_larger_M}, the large momentum transfers - corresponding to large scattering angles - are disfavored by effects due to the finite nuclear size accounted for by means of the nuclear form factor (Sect.\ref{Finite_Nucl_size}).~For instance, in Fig.~\ref{fig:RatiodEdx} the ratios of nuclear stopping powers of electrons in silicon are shown as a function of the kinetic energies of electrons from 200\,keV up to 1\,TeV.~These ratios are the nuclear stopping powers calculated neglecting i) nuclear size effects (i.e., for $\left|F_{\rm exp}\right|^2=1$) and ii) effects due to the finite rest mass of the target nucleus [i.e., in Eq.~(\ref{de/dx_nuclear_el_Nucl_McF}) replacing ${d\sigma^{\rm McF}_{{\rm sc},F,\rm CoM}(T)}/{dT}$ with ${d\sigma^{\rm McF}_{{\rm sc},F}(T)}/{dT}$ from Eq.~(\ref{Factorization_McF_screened_NFF_T})] both divided by that one obtained using Eq.~(\ref{de/dx_nuclear_el_Nucl_McF}).~Above a few tens of MeV, a larger stopping power is found assuming $\left|F_{\rm exp}\right|^2=1$ and, in addition, above a few hundreds of MeV the stopping power largely decreases when effects due to the finite nuclear rest mass are not accounted for.
\par
In Fig.~\ref{fig:el_dEdx} , the nuclear stopping powers in $^7$Li, $^{12}$C, $^{28}$Si and $^{56}$Fe are shown as a function of the kinetic energy of electrons from 200\,keV up to 1\,TeV.~These nuclear stopping powers in MeV\,cm$^2$/g are calculated from Eq.~(\ref{de/dx_nuclear_el_Nucl_McF}) and divided by the density of the medium.~The flattening of the high energy behavior of the curves is mostly due to the nuclear form factor which prevents the stopping power to increase with increasing $T_{\rm max}$.~As expected, the stopping power are slightly (not exceeding a few percent) varied at large energies replacing $F_{\rm exp}$ with $F_{\rm gau}$ or $F_{\rm u}$ (Sect.~\ref{Finite_Nucl_size}).~However, a further study is needed to determine a most suited parametrization of the nuclear form factor\cite{Nagarajan,Duda,Jentschura} particularly for high-$Z$ materials; for instance, in lead the stopping power results to be depressed at energies of about (20-40)\,MeV, while in medium and light nuclei this occurs at energies of the order or above 100\,MeV.% with respect to that calculated at 100\,MeV.
%%%%%%%%%%%%%%%%%%%%%%%%%%%%%%%%%%%%%%%%%%%%%%%%%%%%%%%%%%%%%%%%%%%%%%%%%%%%%%%%%%%%%%%%%%%%%%%%%%%%%%%%%%%%%%%%%%
\begin{figure}[t]
%\vspace{-0.7cm}
\begin{center}
\mbox{%\hspace{-0.3cm}
\psfig{file=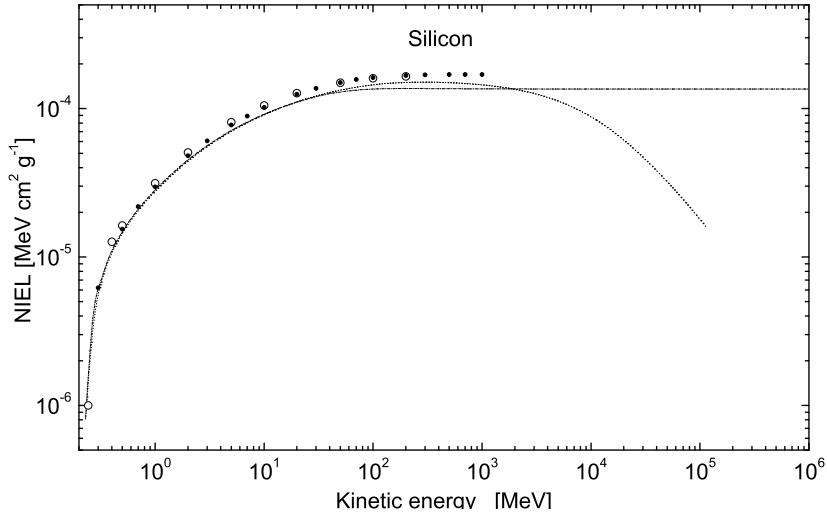,width=4.3in}}
\end{center}
\vspace{-0.2cm}
\caption{For $T_d = 21\,$eV, NIEL (in MeV\,cm$^2$\,g$^{-1}$) in silicon calculated using Eq.~(\ref{eq:NIEL_McF})
as a function of the kinetic energy from 220\,keV up to 1\,TeV (dashed and dotted curve); NIEL values from Messenger et al.\cite{Summers1999} ($\circ$) and Jun et al.\cite{Jun_2009} ($\bullet$) calculated in the laboratory system without accounting for the effects due to the screened Coulomb potential, finite size and rest mass of recoil silicon; the dotted curve is obtained replacing ${d\sigma^{\rm McF}_{{\rm sc},F,\rm CoM}(T)}/{dT}$ with ${d\sigma^{\rm McF}}(T)/{dT}$ [Eq.~(\ref{eq:McF_T})] in Eq.~(\ref{eq:NIEL_McF}).~}
\label{fig:NIEL_elec_silicon}
\end{figure}
%%%%%%%%%%%%%%%%%%%%%%%%%%%%%%%%%%%%%%%%%%%%%%%%%%%%%%%%%%%%%%%%%%%%%%%%%%%%%%%%%%%%%%%%%%%%%%%%%%%%%%%%%%%%%%%%%%%%
%
\section{Non-Ionizing Energy-Loss of Electrons}
\label{El_NIEL}
A relevant process - which causes permanent damage to the silicon bulk structure - is the so-called \textit{displacement damage} (e.g.,~see Chapter~4 of Ref.\cite{LR_3rd}, Refs.\cite{rop_si,Boschini_2011,NIMB2006} and references therein).~Displacement damage may be inflicted when a \textit{primary knocked-on atom} (PKA) is generated.~The interstitial atom and relative vacancy are termed Frenkel-pair (FP).~In turn, the displaced atom may have sufficient energy to migrate inside the lattice and - by further collisions - can displace other atoms as in a collision cascade.~This displacement process modifies the bulk characteristics of the device and causes its degradation.~The total number of FPs can be estimated calculating the energy density deposited from displacement processes.~In turn, this energy density is related to the \textit{non-ionizing energy loss} (NIEL),~i.e., the energy per unit path lost by the incident particle due to displacement processes.
\par
In case of Coulomb scattering of electrons on nuclei, the non-ionizing energy-loss can be calculated using (as discussed in Sect.~\ref{UNscreened_Sect}--\ref{El_Nucl_dE/dx}) the MDCRS or its approximate expression McFDCS [e.g., Eqs.~(\ref{Factorization_Mott_screened_NFF_CoM_T},~\ref{Factorization_McF_screened_NFF_CoM_T}), respectively], once the screened Coulomb fields, finite sizes and rest masses of nuclei are accounted for,~i.e., in MeV/cm
\begin{equation}\label{eq:NIEL_Mott}
- \left(\frac{dE}{dx}\right)_{\rm n,Mott}^{\rm NIEL}  = n_A\,\int^{T_{\rm max}}_{T_d} \!T\,L(T)\,\frac{d\sigma^{\rm Mott}_{{\rm sc},F,\rm CoM}(T)}{dT}\,dT\
\end{equation}
or
\begin{equation}\label{eq:NIEL_McF}
- \left(\frac{dE}{dx}\right)_{\rm n,McF}^{\rm NIEL}  = n_A\,\int^{T_{\rm max}}_{T_d} \!T\,L(T)\,\frac{d\sigma^{\rm McF}_{{\rm sc},F,\rm CoM}(T)}{dT}\,dT\
\end{equation}
[e.g.,~see Equation~(4.113) at page~402 and, in addition, Sections~4.2.1--4.2.1.2 of Ref.\cite{LR_3rd}], where $T$ is the kinetic energy transferred to the target nucleus, $L(T)$ is the fraction of $T$ deposited by means of displacement processes.~The \textit{Lindhard partition function}, $L(T)$, can be approximated using the so-called \textit{Norgett--Robintson--Torrens expression} [e.g.,~see Refs.\cite{Jun2001,Summers2003} and/or Equations~(4.121,~4.123) at pages~404 and~405, respectively, of Ref.\cite{LR_3rd} (see also references therein)].~$T_{\rm de}= T\, L(T)$ is the so-called \textit{damage energy},~i.e.,~the energy deposited by a recoil nucleus with kinetic energy $T$ via displacement damages inside the medium.~In Eqs.~(\ref{eq:NIEL_Mott},~\ref{eq:NIEL_McF}) the integral is computed from the minimum energy $T_d$  - the so-called \textit{threshold energy for displacement},~i.e., that energy necessary to displace the atom from its lattice position - up to the maximum energy $T_{\rm max}$ that can be transferred during a single collision process.~For instance, $T_d$ is about 21\,eV in silicon (e.g., see Table~1 in Ref.\cite{Jun} and references therein) requiring electrons with kinetic energies above $\approx 220\,$keV [e.g.,~see Equation~(4.142) at page~412 of~Ref.\cite{LR_3rd}].
% and 25\,eV in lead (e.g., see Table~22 at page~83 in Ref.\cite{PB_Td} and references therein) requiring electrons with kinetic energies above $\approx 1.12\,$Mev.~
\par
As already discussed with respect to nuclear stopping powers in Sect.~\ref{El_Nucl_dE/dx}, the large momentum transfers (corresponding to large scattering angles) are disfavored by effects due to the finite nuclear size accounted for by the nuclear form factor.~For instance, in Fig.~\ref{fig:RatioNIEL} the ratios of NIELs for electrons in silicon are shown as a function of the kinetic energy of electrons from 220\,keV up to 1\,TeV.~These ratios are the NIELs calculated neglecting i) nuclear size effects (i.e.,~for $\left|F_{\rm exp}\right|^2=1$) and ii) effects due to the finite rest mass of the target nucleus [i.e.,~in Eq.~(\ref{eq:NIEL_McF}) replacing ${d\sigma^{\rm McF}_{{\rm sc},F,\rm CoM}(T)}/{dT}$ with ${d\sigma^{\rm McF}_{{\rm sc},F}(T)}/{dT}$ from Eq.~(\ref{Factorization_McF_screened_NFF_T})] both divided by that one (Fig.~\ref{fig:NIEL_elec_silicon}) obtained using Eq.~(\ref{eq:NIEL_McF}).~Above $\approx 10\, $MeV, the NIEL is $\approx 20\%$ larger assuming $\left|F_{\rm exp}\right|^2=1$ and, in addition, above (100--200)\,MeV the calculated NIEL largely decreases when the effects of nuclear rest mass are not accounted for.~Finally, it has to be remarked that similar results can be obtained neglecting the screening factor: already at energies lower that 200\,keV, $T_d \approx  21\,$eV is much larger than $T_{\rm max}A_{\rm s}$.~
\par
In Fig.~\ref{fig:NIEL_elec_silicon}, %(Fig.~\ref{fig:NIELdEdx_Pb})
for $T_d = 21\,$eV the non-ionizing energy loss (in MeV\,cm$^2$\,g$^{-1}$) calculated using Eq.~(\ref{eq:NIEL_McF}) in silicon %(lead)
is shown (dashed and dotted curve) as a function of the kinetic energy from 220\,keV up to 1\,TeV and is compared with that one tabulated by Messenger et al.\cite{Summers1999} (Jun et al.\cite{Jun_2009}) from $\approx 240\,$keV up to 200\,MeV (1\,GeV).~For the laboratory system, Messenger et al.\cite{Summers1999} used the approximate MDCS found by Doggett-Spencer\cite{Doddett} and Lindhard's partition function numerically obtained by Doran\cite{Doran} without accounting for the effects due to screened Coulomb potential (i.e.,~$\mathfrak{F}^2=1$), finite size (i.e.,~$F^2=1$) and finite rest mass of the silicon target; while, Jun et al.\cite{Jun_2009} followed the approach discussed in Ref.\cite{Lijian} (see the treatment in Sect.~\ref{Mott_R_approx}) to determine an approximate expression of the MDCS and dealt the Lindhard partition function using the modified Norgett--Robintson--Torrens expression found\footnote[2]{Jun et al.~(2009) determined that the usage of the Norgett--Robintson--Torrens expression or, alternatively, the one modified by Akkerman and Barak~(2006) yields similar NIEL values.} by Akkerman and Barak~(2006).~The dotted curve is obtained replacing ${d\sigma^{\rm McF}_{{\rm sc},F,\rm CoM}(T)}/{dT}$ with ${d\sigma^{\rm McF}}(T)/{dT}$ [Eq.~(\ref{eq:McF_T})] in Eq.~(\ref{eq:NIEL_McF}): at 100\,MeV--1\,GeV, the agreement between the latter calculation and values from Messenger et al.\cite{Summers1999} and Jun et al.\cite{Jun_2009} is within several percents.~It has to be remarked (see also Fig.~\ref{fig:RatioNIEL}) that i) above (100--200)\,MeV effects due to screened Coulomb potentials, finite sizes and finite rest masses of nuclei have to be taken into account and ii) for energies between $\approx 100$\,MeV and $\approx1$\,GeV the effects of neglecting the nuclear form factor and finite rest mass of nuclei almost compensate each other.
\section{Conclusions}
\label{Summry_RESULTS}
The treatment of electron--nucleus interactions accounting for effects due to screened Coulomb potentials, finite sizes and finite rest masses of nuclei allows one to determine both the total and differential cross sections, thus, to calculate the resulting nuclear and non-ionizing stopping powers from low (about 200\,keV) up to very high energy (1\,TeV).~
\par
Above a few hundreds of MeV, neglecting the effects of finite rest masses of recoil nuclei, the stopping power and NIEL result to be largely underestimated.~Above a few tens of MeV the finite size of the nuclear target prevents a further increase of both stopping power and NIEL, which approach almost constant values.~The flattening of the high energy behavior of the nuclear and non-ionizing energy-losses is mostly due to the nuclear form factor which prevents stopping powers to increase with increasing $T_{\rm max}$.~However, a further study is needed to determine a most suited parametrization of the nuclear form factor able to provide a satisfactory trend in the energy region below about hundred MeV also for high-$Z$ materials.
\par
Finally, at 100\,MeV--1\,GeV an agreement to within several percents was obtained between the present calculation with respect to the NIEL values from Messenger et al.\cite{Summers1999} and Jun et al.\cite{Jun_2009}.\,.
\bibliographystyle{ws-procs9x6}
\bibliography{ws-pro-sample}

\begin{thebibliography}{99}

\bibitem{Meyer_V}
P. Meyer and R. Vogt,~\textit{Phys. Rev. Lett.} 8 (1962), 387--389.

\bibitem{Owens2010}
M.J. Owens, T.S. Horbury and C.N. Arge,~\textit{Astrophys. J.} 714 (2010), 1617, doi: 10.1088/0004-637X/714/2/1617.

\bibitem{LR_3rd}
C. Leroy and P.G. Rancoita, \textit{Principles of Radiation Interaction in Matter and Detection}, 3rd Edition, World Scientific (Singapore) 2011.

\bibitem{rop_si}
C. Leroy and P.G. Rancoita, Particle Interaction and Displacement Damage in Silicon Devices operated in Radiation Environments, \textit{Rep. Prog. in Phys.} 70 (no. 4)(2007), 403--625, doi: 10.1088/0034-4885/70/4/R01.

\bibitem{Boschini}
M. Boschini et al.,~Geant4-based application development for NIEL calculation in the Space Radiation Environment,
Proc. of the 11th ICATPP, October 5--9 2009, Villa Olmo, Como, Italy, C. Leroy, P.G. Rancoita, M. Barone, E.
Gaddi, L. Price and R. Ruchti Editors, World Scientific, Singapore(2010), 698--708, IBSN: 10-981-4307-51-3.

\bibitem{Boschini_2011}
M. Boschini et al.,~Nuclear and Non-Ionizing Energy-Loss for Coulomb Scattered Particles from Low Energy up to Relativistic Regime in Space Radiation Environment, Proc. of the 12th ICATPP, October 7--8 2010, Villa Olmo, Como, Italy, S. Giani, C. Leroy and P.G. Rancoita, Editors, World Scientific, Singapore (2011), 9--23, IBSN: 978-981-4329-02-6.

\bibitem{geant4}
S. Agostinelli et al.,~Geant4 – a simulation toolkit, \textit{Nucl. Instr. and Meth. in
Phys. Res. A} 506~(2003), 250--303. \\ See also the web site:
\textit{http://geant4.web.cern.ch/geant4/}

\bibitem{Mott1}
N.F. Mott,~\textit{Proc. Roy. Soc. A} 124~(1929), 425--442; \textit{A} 135 (1932), 429--458.

\bibitem{Mott_book}
N.F. Mott and H.S.W. Massey, The Theory of Atomic Collisions - 3rd Edition-~(1965), Oxford University Press, London.

\bibitem{Wentzel}
G. Wentzel,~\textit{Z. Phys.} 40~(1927), 590--593.

\bibitem{Born1}
M. Born,~\textit{Z. Phys.} 38~(1926), 803.

\bibitem{Idoeta}
R. Idoeta and F. Legarda,~\textit{Nucl. Instr. and Meth. in Phys. Res. B} 71~(1992), 116--125.

\bibitem{Berger_comp}
 M.J. Berger,~Monte Carlo Calculation of the Penetretion and Diffusion of Fast Charged Particles, in Methods in Computational Physics \textbf{vol. 1}~(1963), B. Alder, S. Fernbach and M. Rotenberg Editors, Acdemic Press, New York, 135--215.

\bibitem{Fernandez1}
J.M. Fernandez-Vera, R. Mayol and F. Salvat, \textit{Nucl. Instr. and Meth. in Phys. Res. B} 82, (1993) 39--45.

\bibitem{Bartlett_Watson}
J.H. Bartlett and R.E. Watson,~\textit{Proc. Am. Acad. Arts Sci.} 74~(1940), 53.

\bibitem{Sherman}
N. Sherman,~\textit{Phys. Rev.} 103~(1956), 1601--1607.

\bibitem{McKinley}
A.Jr. McKinley and H. Feshbach,~\textit{Phys. Rev.} 74~(1948), 1759--1763.

\bibitem{Curr}
R.M. Curr,~\textit{Proc. Phys. Soc. }(London) A68~(1955),
156--164.

\bibitem{Cahn}
J.H. Cahn, \textit{J. of Appl. Phys.} 30 (1959), 1310--1316.

\bibitem{Feshbach1}
H. Feshbach,~\textit{Phys. Rev.} 88~(1952), 295--297.

\bibitem{Doddett}
J.A. Doggett and L.V. Spencer,~\textit{Phys. Rev.} 103~(1956), 1597–-1601.

\bibitem{Lijian}
T. Lijian, H. Quing and L. Zhengming,~\textit{Radiat. Phys. Chem.}
45~(1995), 235--245.

\bibitem{Oen}
O.S. Oen, \textit{Nucl. Instr. and Meth. in Phys. Res. B} 33, (1988) 744--747.

\bibitem{Seitz}
F. Seitz and J.S. Koehler,~\textit{Solid State Physics}
\textbf{vol. 2}, edited by F. Seitz and D. Turnbull, Academic
Press Inc., New York~(1956).

\bibitem{Fernandez}
J.M. Fernandez-Vera et al.,~\textit{Nucl. Instr. and Meth. in Phys. Res. B} 73~(1993), 447--473.

\bibitem{Moliere}
von G. Moli\`{e}re,~\textit{Z. Naturforsh.} A2~(1947), 133--145; A3 (1948), 78--97.

\bibitem{m_sca_other}
H.A. Bethe,~\textit{Phys. Rev.} 89~(1953), 1256--1266.

\bibitem{Butkevick}
A.V. Butkevick,~\textit{Nucl. Instr. and Meth. in Phys. Res. A} 488~(2002), 282--294.

\bibitem{Thomas}
L.H. Thomas,~\textit{Proc. Cambridge Phil. Soc.} 23 (1927) , 542.

\bibitem{Fermi_TF}
E. Fermi,~\textit{Z. Phys.} 48~(1928), 73--79.

\bibitem{Zeitler}
E. Zeitler and A. Olsen,~\textit{Phys. Rev.} 136~(1956), A1546–-A1552.

\bibitem{Helm}
R.H. Helm,~\textit{Phys. Rev.} 104~(1956), 1466–-1475.

\bibitem{Ho57}
R. Hofstadter,~\textit{Ann. Rev. Nucl. Sci.} 7~(1957), 231.

\bibitem{DeVries}
H. De Vries, C.W. De Jager, and C. De Vries,~\textit{Atomic Data and Nuclear Data Tables}
\textbf{36} (1987), 495.

\bibitem{Nagarajan}
M.A. Nagarajan and L. Wang,~\textit{Phys. Rev. C} 10~(1974), 2206–-2209.

\bibitem{Duda}
G. Duda, A. Kemper and P. Gondolo, \textit{J. Cosm. Astrop. Phys.} 04 (2007), 012, doi:10.1088/1475-7516/2007/04/012

\bibitem{Jentschura}
U.D. Jentschura and V.G. Serbo, \textit{E. Phys. J. C} 64 (2009), 309--317.

\bibitem{NIMB2006}
C. Consolandi et al.,~\textit{Nucl. Instr. and Meth. in Phys. Res. B} 252~(2006), 276.

\bibitem{Jun2001}
I. Jun, \textit{IEEE Trans. on Nucl. Sci.} 48 (2001), 162--175

\bibitem{Summers2003}
S.R. Messenger et al.,~\textit{IEEE Trans. on Nucl. Sci.} 50 (2003), 1919--1923.

\bibitem{Jun}
I. Jun, M.A Xapsos, S.R. Messenger, E.A. Burke, R.J. Walters, G.P. Summers and T. Jordan,~\textit{IEEE Trans. on Nucl. Sci.}
50~(2003), 1924--1928.

%\bibitem{PB_Td}
%G.S. Was,~\textit{Fundamentals of radiation materials science: metals and alloys} Springer (Berlin), 2007.

\bibitem{Summers1999}
S.R. Messenger et al.,~\textit{IEEE Trans. on Nucl. Sci.} vol. 46, no. 6 (1999), 1595-1601.

\bibitem{Jun_2009}
I. Jun, W. Kim and R. Evans,~\textit{IEEE Trans. on Nucl. Sci.} vol. 56~(2009), 3229--3235.

\bibitem{Akkerman_2006}
A. Akkerman and J. Barak,~\textit{IEEE Trans. on Nucl. Sci.} vol. 53~(2006), 3667--3674.

\bibitem{Doran}
D.G. Doran,~\textit{Nucl. Sci. Eng.} 49 (1972), 130--144.


\end{thebibliography}

\end{document}